# `Optimization of Higher-Order Harmonic Surface Tessellations for Additively Manufactured Air-to-Air Heat Exchangers

*Patrick Adegbaye[a], Aigbe E. Awenlimobor[a], Justin An[a], Zhang Xiao[a], Jiajun Xu[a*]*

[a] *Department of Mechanical Engineering, University of the District of Columbia, Washington, DC, USA*

*Corresponding authors: E-mail addresses: jiajun.xu@udc.edu (J. Xu)*

## Abstract

Air-to-air heat exchangers are vital for energy recovery and thermal management but often suffer from reduced effectiveness, high pressure losses, and increased pumping power in conventional designs. Advances in additive manufacturing have enabled nature-inspired geometries, such as lattice and triply periodic minimal surface (TPMS) structures, which enhance heat transfer through complex first-order surfaces but frequently cause excessive pressure drops. This study proposes an optimized higher-order harmonic heat-transfer surface tessellation developed through an optimization framework integrating analytical and numerical methods. The goal is to improve the overall thermal-hydraulic performance of the heat exchanger over a range of operating conditions. Results of sensitivity analysis show that secondary surface modification of this type can yield significant increase in the effectiveness reaching up to 70% although with associated increase in the pressure drop. The secondary surface wave frequency was found to be a more important control parameter than the amplitude in achieving high thermal-hydraulic performance. Additionally, we show that the optimized second order harmonic-type structure achieved relatively higher effectiveness and lower pressure-drop than the gyroid structure in the turbulent flow regime for $Re \geq 7000$. Although the gyroid TPMS structure had relatively higher effectiveness in the laminar and weakly turbulent flow regime, the associated pressure drop was found to be significantly higher than that of the harmonic-type structure. While the gyroid structure had significantly higher heat transfer efficiency with an average Nusselt number about 5.2 times that the harmonic-type structure, the corresponding flow resistance was likewise observed to be significantly higher on average, about 8.4 times higher in the laminar flow regime and about 1.87 times higher in the turbulent flow regime. The London area goodness factor (AGF) was used to assess the overall thermal-hydraulic performance of both architectures which showed that the harmonic-type structure performed better in the laminar flow regime ($Re \leq 2000$) which had an average AGF of about 1.72 times the AGF of the gyroid structure and reaching up 2.18 times the gyroid's AGF when $Re = 500$. However, the gyroid structure outperformed the harmonic-type structure in the turbulent flow regime. The findings from this study demonstrate a balanced pathway for additively manufacturable, high-performance air-to-air heat exchangers, offering compact, energy-efficient solutions for applications in building ventilation, aerospace, and electronics cooling.

## Introduction

An air-to-air heat exchanger (AAHX) is a specialized device engineered to facilitate the transfer of thermal energy between two separate air streams without permitting them to mix. This separation is critical in maintaining air quality and preventing cross-contamination, especially in environments where hygiene and air purity are paramount [1]. The design and development of air-to-air heat exchangers have evolved considerably over the years, leading to significant improvements in both performance and versatility. Modern AAHX development is commonly formulated as a coupled thermo-hydraulic optimization problem to maximize heat transfer while constraining pressure drop (Δp), size, and manufacturability. Kim et al. (2022) expanded on such methodologies when exploring compact cross-/counter-flow energy recovery ventilator (ERV) optimization methodologies [2]. Non-traditional core geometries to expand the achievable heat transfer-pressure drop tradeoff have also been explored in parallel, including multi-morphological TPMS-based microchannel concepts aimed at thermal performance optimization [3].

The gyroid lattice structure is a complex, three-dimensional structure characterized by continuous, periodic, and non-self-intersecting channels. They have become popular due to being able to contain a large surface area within a small volume and can be described with few geometric parameters. Bartlett et al. (2024) describes TPMS geometries as periodic minimal surfaces that split space into intertwined, non-intersecting channels, which gives high surface-area-to-volume ratio and avoids the sharp edges found in many compact-fin designs [4]. These designs are especially effective in addressing the limitations of traditional heat exchanger surfaces, which often suffer from low surface area and inefficient heat transfer [5, 6]. By enabling increased surface contact between the air streams and the exchanger walls, TPMS geometries significantly enhance thermal performance. Building on this, Tang, Chen, and Zhao (2024) show that gradient and field-synergy-based optimization strategies can further improve performance by reshaping local flow-thermal-gradient alignment to increase heat transfer at a given flow penalty [7].

Gyroid structures offer a balance between mechanical strength and fluid dynamics, making it suitable for a wide range of applications. However, the benefits of these advanced geometries are most pronounced under single-phase airflow conditions. In relation, Hao et al. (2023) developed convective heat transfer correlations for TPMS-based heat exchangers, formalizing Reynolds number (Re) and Nusselt number (Nu) relationships across representative TPMS families to support performance prediction and design space exploration [8]. When dealing with multiphase or turbulent flows, the performance gains may diminish due to complex interactions between the air streams and the surface topology. Chen, Luo, and Wei (2024) expands that TPMS-induced secondary flows and inter-channel interactions can increase Δp and alter mixing behavior in ways that reduce net performance gains outside targeted operating envelopes, particularly as flow regimes and transport mechanisms change [9].

Beyond thermal-hydraulic performance, TPMS cores are attractive for lightweighting and multifunctionality because the same periodic architecture can be co-optimized for stiffness, mass, and heat transfer, aligning with broader trends in aerospace/transport heat exchanger lightweight design. Liu et al. (2024) discusses how these benefits are increasingly pursued through additive manufacturing (AM), which enables fabrication of complex gyroid/TPMS lattices and two-fluid exchange topologies that are difficult or impossible to produce conventionally [10]. Careri, Zocca, and Baldini (2023) emphasize the value of AM-enabled compactness and mass reduction for heat exchanger applications [11], while Meyer et al. (2025) further highlight the performance leverage available through deliberate internal-structure design in additively manufactured heat exchangers [12].

Manufacturing these intricate structures presents several challenges. Practical concerns include holding tight dimensions, limiting unwanted porosity, keeping the two air circuits leak-tight, and understanding how as-built roughness changes friction and heat transfer. Feng, Xu, and Liu (2024) address some of these issues through hierarchical TPMS lattice designs and optimization strategies that consider performance and structure together [13]. Additive manufacturing techniques, such as selective laser sintering or stereolithography, have shown promise in fabricating these geometries, but scalability and cost remain barriers to widespread adoption. From an optimization standpoint, Kobayashi et al. (2021) reinforce the idea that manufacturability constraints need to be built into the topology design process for two-fluid heat exchange, not treated as an afterthought [14]. In addition, Nazir et al. (2019) summarize the broader state of cellular-structure additive manufacturing, noting recurring challenges with repeatability, defect control, and cost-effective scaling [15].

Despite the progress made in geometric optimization, most existing research on heat exchangers tends to focus on a single operational parameter. Simulations often assume steady-state or laminar airflow conditions, which simplifies the analysis but fails to capture the complexities of real-world environments. Waters (2023) explains how optimization and additive manufacturing can produce highly efficient three-dimensional surface structures, but the broader literature still commonly relies on steady-state and/or laminar-flow assumptions to simplify the analysis [16]. In practice, airflow is rarely uniform; it fluctuates due to changes in occupancy, equipment usage, and external weather conditions. These dynamic factors introduce turbulence, pressure variations, and temperature gradients that can significantly impact heat exchanger performance.

Because of this, gains reported under idealized simulations do not always translate into the same benefit in practice. One clear real-world complication is flow imbalance and maldistribution. Kaminski et al. (2024) specifically compare counterflow air-to-air heat exchangers under unbalanced flow conditions and show that off-design operation can noticeably change heat transfer efficiency relative to balanced, uniform assumptions [17]. Such results reinforce why simplified performance metrics can be misleading, and why more realistic modeling is increasingly important.

To address these limitations, there is a growing opportunity for research to move beyond geometric modifications and explore alternative strategies for enhancing heat exchanger functionality. For example, Liu et al. (2023) combines numerical and experimental work on additively manufactured TPMS heat exchangers and demonstrate that the thermal and hydraulic response is strongly tied to geometry, manufacturing condition, and operations, highlighting the need to validate predicted results beyond idealized assumptions [18]. These include the use of advanced materials with tailored thermal properties, integration of active control systems, and development of hybrid configurations that combine multiple heat transfer mechanisms [19]. Additionally, machine learning and artificial intelligence can be leveraged to analyze large datasets and identify optimal design parameters under varying conditions. Guo, Lin, and Zheng (2024) use machine learning to design and optimize staggered fin structures, showing how data-driven models can speed up optimization and help identify robust parameter choices across broader conditions than a single simulation process might cover [20].

In response to these gaps, this study introduces a novel approach to heat exchanger design through the development of an optimized, higher-order harmonic heat transfer surface tessellation. This innovative geometry is generated using a multi-objective optimization framework that combines analytical modeling with numerical simulations. The framework considers multiple performance metrics, including thermal efficiency, flow resistance, manufacturability, and structural integrity. By balancing these objectives, the resulting design offers improved heat transfer capabilities while minimizing pressure drop and material usage. The proposed tessellation is specifically tailored for air-to-air applications, where maintaining airflow continuity and minimizing energy loss are critical. Its harmonic structure promotes uniform distribution of air across the surface, reducing hotspots and enhancing overall thermal performance. Furthermore, the design is compatible with modern manufacturing techniques, enabling scalable production and integration into existing HVAC systems [21].

## Methodology

The representative computational domain considers a unit cell of the HX structure which was constructed in COMSOL Multiphysics (COMSOL Inc., Natick, Massachusetts, USA) with ideally repeated units in the lateral directions perpendicular to the flow direction as shown in Figure 1(a). The geometry of the heat transfer surface area (cf. Figure 1(b)) was defined analytically as a second-order harmonic function given as

$$z = A\,sin(n\pi x) + B\,cos(m\pi y)\,cos(m\pi x)\,e^{-k\sqrt{x^2+y^2}} + \epsilon \qquad (1)$$

In this expression, (A) and (B) are amplitude coefficients, (n) and (m) are spatial frequency parameters, and (k) introduces localized undulations that form controlled surface dimples and ridges. These features are intentionally designed to generate small-scale vortices and promote mixing within the airflow, thereby increasing the residence time near the heated surface. A length of 4cm and fin thickness of 2 mm was assumed for the unit cell. Unlike more complex geometries such as Triply Periodic Minimal Surfaces (e.g., gyroid or diamond structures), this approach aims to attain similar thermal enhancement through simpler, more manufacturable surface modulations.

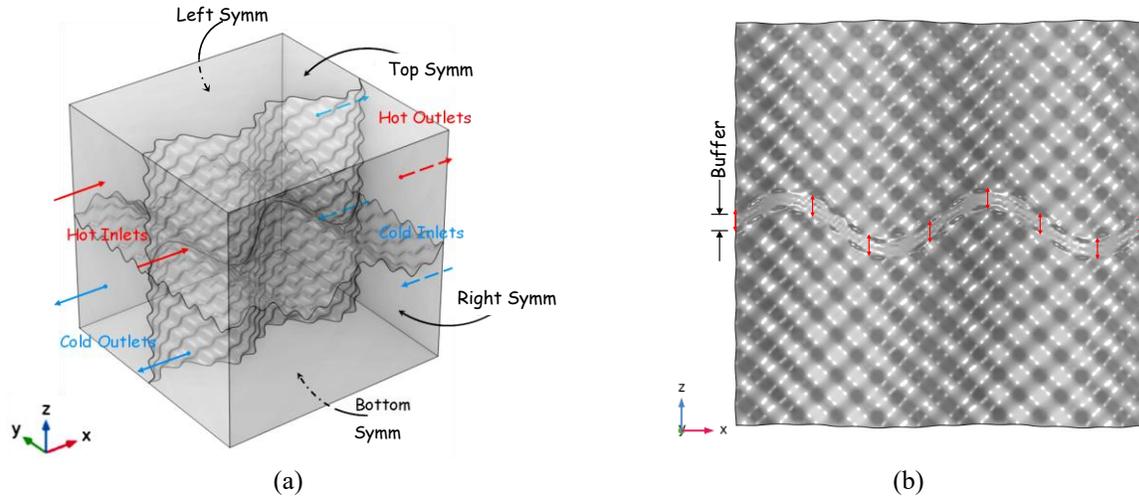

(a) (b)

**Figure 1:** (a) representative unit cell of the tessellated wavy channel HX architecture, (b) side view of the tessellated fin surface showing the harmonic features and buffer zone at the fin-fin intersection.

The gyroid structure was obtained from COMSOL's inbuilt CAD library and an envelope dimension and fin thickness similar to the tessellated wavy channel architecture was adopted (cf. Figure 2(a)). The gyroid TPMS interior fin parametric surface (cf. Figure 2(b) can be defined by the implicit equation given as [22]

$$sin\left(\frac{2\pi}{L}x\right)cos\left(\frac{2\pi}{L}y\right) + sin\left(\frac{2\pi}{L}y\right)cos\left(\frac{2\pi}{L}z\right) + sin\left(\frac{2\pi}{L}z\right)cos\left(\frac{2\pi}{L}x\right) \leq c \qquad (2)$$

Irrelevant edges on either side of the fin surface that were internally generated by the CAD library during model import were removed to ensure a smooth surface texture and to avoid any unnecessary feature refinement during domain discretization that do contribute to defining its form which ensures a uniform mesh distribution.

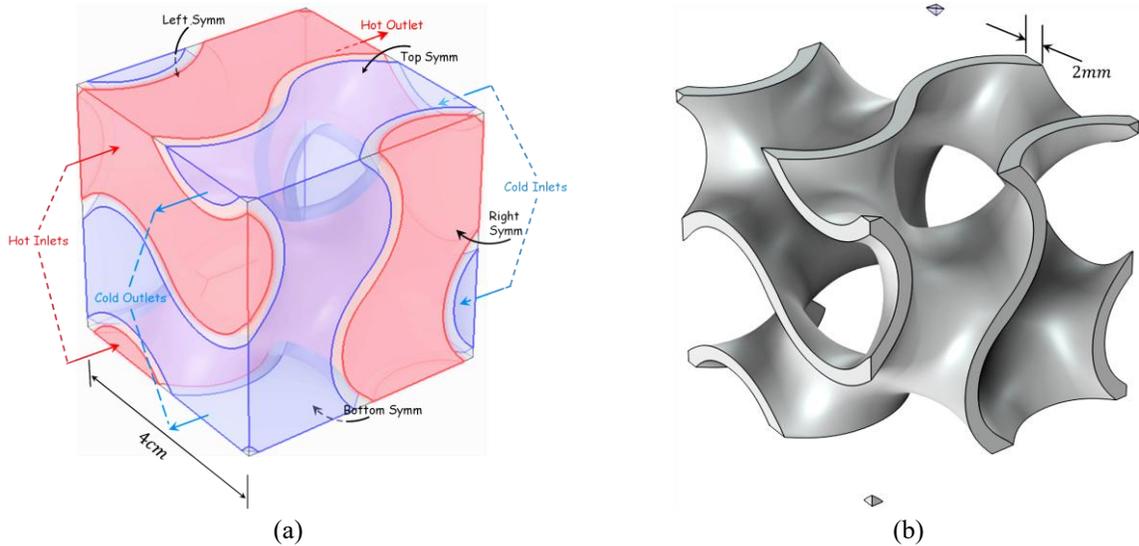

(a) (b)

**Figure 2:** (a) representative unit cell of the gyroid HX architecture obtained from COMSOL CAD library, (b) isolated view of the convoluted fin structure separating the hot and cold fluid zones and showing the physical thickness.

### FEA Model Development:

The computational domain of the unit cell structure is shown in Figure 1(a) for the tessellated wave structure and Figure 2(a) for the gyroid structure counterpart. Air flows into the hot inlet at a velocity of $\bar{u}$ m/s and temperature of

$\mathcal{T}_h^i$ [K] and into the cold inlet at the same velocity and temperature of $\mathcal{T}_c^i$ [K] while atmospheric pressure conditions are defined at the flow outlets. The flow is assumed to be fully developed neglecting flow entrance and exit effects and considering the minimum entrance length for the flow regime considered is much greater than the characteristics flow length of the unit cell, $L_c$, i.e. $L_c/D_h \ll \min(0.05 Re_{lam}, 4.4 Re_{tur}^{1/6})$. We consider symmetry conditions on the bounding surfaces perpendicular to the flow direction [26,29]. Because the fin thickness is relatively small compared to the physical dimensions of the unit cell and due to the tessellated topology of the interior surface which would result in excessively distorted geometry or very fine mesh requirement if the thickness of were to be fully resolved, a thermally thin layer approximation with a shell thickness of 2 mm is used to define the fin structure. Temperature and pressure dependent properties of air are considered for the fluid medium in the simulation using inbuilt material libraries from the COMSOL software, and aluminum material is considered for the solid fin structure as in typical applications due to its weight reduction and high thermal conductivity.

The governing equations for mass, and momentum conservation and energy balance describing the conjugate heat transfer (CHT) and high-Reynolds-number turbulent fluid flow are based on the shear stress (SST) Reynolds-Averaged Navier-Stokes (RANS) $\kappa - \omega$ equations as given in equations (3)-(7)

$$\frac{\partial \overline{\rho u_j}}{\partial x_j} = 0 \tag{3}$$

$$\frac{\partial \bar{\rho} u_i u_j}{\partial x_j} = \frac{\partial P}{\partial x_i} + \frac{\partial}{\partial x_j}\left(\mu \frac{\partial (u_i)}{\partial x_j} - \rho \overline{u_i u_j}\right) \tag{4}$$

$$\frac{\partial (\rho u_i T)}{\partial x_i} = \frac{\partial}{\partial x_j}\left(\frac{\kappa}{C_p}\frac{\partial T}{\partial x_i}\right) \tag{5}$$

The CHT model couples conduction within solid structures with convection in the airflow. The turbulent kinetic energy (k) and dissipation rate (ω) equations are given as:

$$\frac{\partial (\rho u_i k)}{\partial x_i} = \frac{\partial}{\partial x_j}\left(\Gamma_k \frac{\partial k}{\partial x_i}\right) + G_k - Y_k + S_k \tag{6}$$

$$\frac{\partial (\rho u_i \omega)}{\partial x_i} = \frac{\partial}{\partial x_j}\left(\Gamma_\omega \frac{\partial \omega}{\partial x_i}\right) + G_\omega - Y_\omega + S_\omega + D_\omega \tag{7}$$

In equations (3)-(7), $\rho$ is the density of air, $\mu$ is the dynamic viscosity, $u_i$ is the flow velocity, T is the temperature $\partial/\partial x_i$ is the gradient operator, $\kappa$ is the thermal conductivity, $C_p$ is the specific heat capacity, $\Gamma_k$ is the Effective diffusivity of k. $G_k$ is the Production of turbulent kinetic energy. $Y_k$ is the Dissipation of turbulent kinetic energy. $S_k$ is the Source term.

For the laminar flow model, only equations (2) – (4) applies without averaging the quantities and the Reynold's stress $-\rho \overline{u_i u_j}$ is discarded.

**Domain Discretization**

The fluid domain is discretized using 10 nodes quadratic tetrahedral serendipity elements with quadratic velocity interpolation and linear pressure (i.e. $P_2+P_1$) formulation and with a biased meshing having a minimum element size of 1/5$^{th}$ the smallest feature dimension (i.e. the amplitude of the secondary feature – B), a maximum element size of 1/5$^{th}$ the size of the envelope (W) and a maximum growth rate of 1.5. Mesh refinement is carried out to resolve the interior tessellated surfaces using a minimum size of 1/5$^{th}$ the amplitude of the secondary surface – B, and a maximum element size of 1/5$^{th}$ the amplitude of the primary surface – A, and a maximum growth rate of 1.35. Additionally, boundary layer refinement with 3 layers and a transition ratio of 1.25 is considered near the interior walls of the channel. Overall, a total of 892,774 elements were obtained with the above settings for the base geometry although this number changes with the surface parameters. Figure 3(a) below shows the resulting discretization of the fluid domain, while the section view in Figure 3(b) provides detail of the mesh resolution on the interior walls.

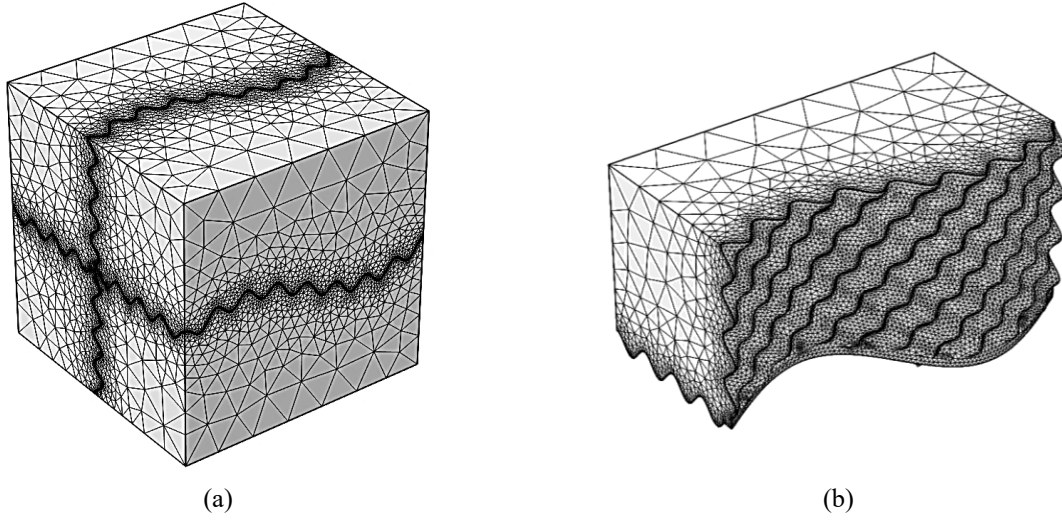

(a) (b)

**Figure 3:** (a) Discretization of the computational domain for the tessellated wavy HX architecture showing inflation layers near the interior walls, (b) single channel view showing the mesh resolution on the tessellated wavy interior surface.

Although COMSOL's in-built solver was set to automatically resolve the first layer height, $y_1$ of the grid in order to accurately capture the laminar boundary layer, we validate the computed $y_1$ based on recommended values of the distance in wall coordinates $y^+$, given as $y^+ = \rho y_1 \bar{u}_\tau / \mu$, where $\bar{u}_\tau$ is the friction velocity computed from $\bar{u}_\tau = \sqrt{0.5 C_f \bar{u}^2}$, and the skin friction coefficient for laminar flow is computed from the Blasius solution given as $C_f = 0.664 Re^{-1/2}$ or from the Schlichting's equation for a broader range of $Re$ such that $Re < 10^9$, given as $C_f = (2 \log_{10} Re - 0.65)^{-2.3}$. The recommended upper bound value of $y^+$ for enhanced wall treatment is $y^+ < 5$ [23]. For the range of laminar flow Reynold's number considered in this work, the minimum upper bound value for $y_1$ was obtained as $y_1 = 0.0593 \ cm$ corresponding to $Re = 2000$, which is far greater than $y_1$ automatically computed by the COMSOL's solver (i.e. $y_1 = 0.016 \ cm$). For the turbulent flow cases, "Standard Wall Functions" was activated in the turbulent model to resolve the thin viscous sublayer which assumes a value of $y^+ > 30$.

Likewise, for the gyroid structure, similar element type formulation and order was used as the tessellated wavy channel architecture for discretizing both the fluid domain and solid fin thickness. For the fluid domain, a minimum element size of 0.25mm and maximum size of 1/5th the size of the envelope (W) with a maximum growth rate of 1.2 was deemed sufficient (cf. Figure 4(a)) while a minimum element size of 0.25mm with a growth rate of 1.15 was used to discretize the solid fin thickness (cf. Figure 4(b)). Boundary layer refinement with 3 layers and a transition ratio of 1.25 is likewise considered near the interior walls of the channel. The resulting mesh yields a total of 479,170 domain elements. In addition, sensitivity study was carried out to ensure mesh sufficiency in accurately predicting the various output responses.

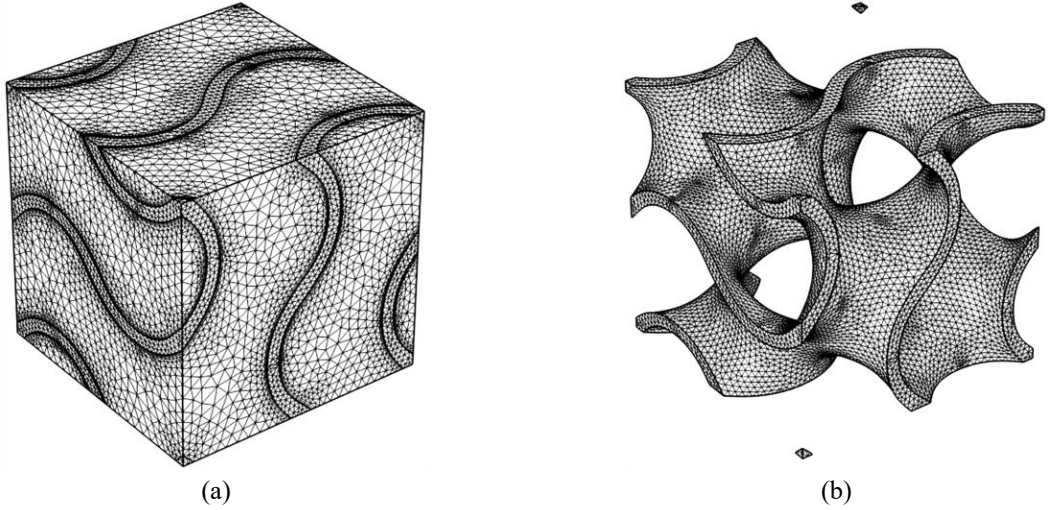

(a) (b)

**Figure 4:** (a) Discretization of the computational domain for the gyroid HX architecture showing mesh refinement near the interior walls, (b) isolated view showing the discretization of the solid fin structure.

**Definition of Model Parameters**

The performance of the heat exchanger is assessed based on the pressure drop, $\Delta P$ and the 'approximate' effectiveness, $\varepsilon$, where $\Delta P$ is defined as the sum of the hot and cold side net difference between the plane average inlet and outlet pressure, while the effectiveness, $\varepsilon$ is given as

$$\varepsilon \approx \frac{\mathcal{T}_c^o - \mathcal{T}_c^i}{\mathcal{T}_h^i - \mathcal{T}_c^i} \qquad (8)$$

Another important quantity used to characterize the heat exchangers performance is the logarithmic mean temperature difference ($\Delta T_m$) defined as

$$\Delta T_m = \frac{\Delta T_1 - \Delta T_2}{\ln(\Delta T_1/\Delta T_2)} \qquad (9)$$

where, $\Delta T_1 = \mathcal{T}_h^i - \mathcal{T}_c^o$, and $\Delta T_2 = \mathcal{T}_h^o - \mathcal{T}_c^i$. With the subscript '*h*' and '*c*' refers to the hot and cold side respectively and superscript '*i*' and '*o*' refer to the inlet and outlet respectively. The average inlet fully developed input flow velocity, $\bar{u}$ used in the simulations is determined from the operating flow regime, which is quantified by the Reynolds number, $Re$ given as:

$$Re = \frac{\rho \bar{u} D_h}{\mu} \qquad (10)$$

where $D_h$ is the hydraulic diameter given as $D_h = 4 V_c/A_{HT}$. $V_c$ is a single channel fluid volume and $A_{HT}$ is its heat transfer surface area assuming full flow, $\rho$ and $\mu$ are the fluid density and dynamic viscosity evaluated at the mean fluid temperature defined as $\mathcal{T}_m = 1/2 \left( \mathcal{T}_c^i + \mathcal{T}_h^i \right)$. Typical expressions for the Nusselt number, $Nu$ for standard compact heat exchanger geometries found in literature would not apply to more complicated lattice structure [Reynold's et al] such as those investigated in this work. We report the Nusselt number, $Nu$ which is a dimensionless representation of the overall heat transfer coefficient, $h$ and develop correlations for $Nu$ over a range of flow conditions by a fitting operation using standard form of the Seider and Tate's equation given as [24]

$$Nu = \frac{hD_h}{\kappa} = a_0 + a_1 \left( Pr \times Re \times \frac{D_h}{L_c} \right)^{1/3} \left( \frac{\mu}{\mu_w} \right)^{0.14} \qquad (11)$$

where, $a_0$, and $a_1$ are fitting parameters obtained from linear regression analysis, $\kappa$ is the bulk fluid thermal conductivity, $Pr = \mu C_p/\kappa$ is the Prandtl number, $C_p$ is the bulk fluid specific heat capacity, with all properties evaluated at the mean fluid temperature, $T_m$. We assume for air, $\mu(T_m) \approx \mu(T_w)$. The heat transfer coefficient, $h$ is computed from the average heat flux through the fin, $\dot{q}$ and the LMTD, $\Delta T_m$ according to expression given as

$$h = \frac{\dot{q}}{\Delta T_m} \tag{12}$$

where $\dot{q} = \dot{Q}/A_{ht}$, $A_{ht}$ is a single flow channel heat transfer surface area, and $\dot{Q}$ is the heat rate given as

$$\dot{Q} = \dot{m}C_{ph}(T_h^o - T_h^i) = -\dot{m}C_{pc}(T_c^o - T_c^i) \tag{13}$$

where $\dot{m} = \rho \bar{u} A$, $A$ is the effective flow channel cross-sectional area. The Seider - Tate's type expression for laminar flow is best suited for our analysis since the Dittus - Boelter or Seider - Tate type expressions for turbulent flow are more applicable for flow regimes with $Re > 10000$ [25]. Moreover, the expression is relatively simple and easy to use in comparison to more complex expressions for the Nusselt number found in literature [25]. The Fanning friction factor $f$, is another dimensionless parameter used to assess the performance of the heat exchanger, which is a dimensionless form of pressure drop across the channel and characterizes the flow resistance [22]. We likewise determine correlations for $f$ in the laminar ($Re \leq 2000$) and turbulent ($Re > 2000$) flow regimes via regression analysis and using expression similar to [22] and given as

$$f \cdot Re = \tau_w \Big/ \frac{1}{2}\rho \bar{u}^2 \cdot Re = \left(\frac{\Delta P}{4}\frac{D_h}{L}\right) \Big/ \frac{1}{2}\rho \bar{u}^2 \cdot Re = b_0 + b_1 Re^m \tag{14}$$

where, $b_0$, $b_1$ and $m$ are fitting parameters obtained from linear regression analysis. A common dimensionless metric for assessing the overall thermal-hydraulic performance of the heat transfer surface is the London Area Goodness factor (AGF) which is the ratio of the Chilton-Colburn factor, $j$ to the Fanning friction factor, $f$ (i.e. $j/f$) where $j$ is given as [26]

$$j = \frac{Nu}{RePr^{1/3}} \tag{15}$$

A high $j/f$ indicates relatively efficient heat transfer surface. Another important overall HX performance criteria is the Thermal Enhancement Factor (TEF) denoted by $\eta$, that represents the heat transfer intensity at equivalent pump power requirement and defined as [27]

$$\eta = \frac{j/j_0}{(f/f_0)^{1/3}} \tag{16}$$

where the subscript, '0' refers to the reference geometry and $\eta > 1$ signifies relatively superior thermal-hydraulic performance and vice versa.

**Material Properties**

The physical properties of the working fluid (air), and the solid fin material (aluminum) are presented in this section. The default materials properties were obtained from the COMSOL material libraries. The properties for the solid aluminum fin are constant values (cf. Table 1), while that of the working fluid (air) are polynomial expressions that depend on the temperature and pressure solution variables to account for compressibility effects. The expressions for the air properties including the density, $\rho$, specific heat capacity, $C_p$, thermal conductivity, $\kappa$, and dynamic viscosity, $\mu$ are given in eqns. (17) – (20) below

$$\rho(p,T) = p/\bar{R} \cdot T \tag{17}$$
$$C_p(T) = 1047.6366 - 0.3726\,T + 9.453 \times 10^{-4}T^2 - 6.0241 \times 10^{-7}T^3 + 1.2859 \times 10^{-10}T^4 \tag{18}$$
$$\kappa(T) = -2.2758 \times 10^{-3} + 1.1548 \times 10^{-4}\,T - 7.9025 \times 10^{-8}T^2 + 4.117 \times 10^{-11}T^3 \\ - 7.4386 \times 10^{-15}T^4 \tag{19}$$
$$\mu(T) = -8.3828 \times 10^{-7} + 8.3572 \times 10^{-8}\,T - 7.6943 \times 10^{-11}T^2 + 4.6437 \times 10^{-14}T^3 \\ - 1.0659 \times 10^{-17}T^4 \tag{20}$$

where $\bar{R} = R/M$, $R$ is the universal gas constant given as $R = 8.31445 \; J/mol \cdot K$ and $M$ is the molar mass of air given as $M = 28.9647 \; g/mol$. Average properties for air evaluated at the mean fluid temperature for the specified HX base operating condition (i.e. at $T_m = 40^0C$) also appear in Table 1 below. The air is assumed to be moist free with negligible relative humidity and obeys standard compressible ideal gas behavior.

**Table 1:** Physical properties of solid HX aluminum fin and the primary working fluid (air) at $T_m = 40^0C$.

| Material | $\rho$ [kg/m³] | $C_p$ [J/kg·K] | $\kappa$ [W/m·K] | $\mu$ [Pa·s] |
|---|---|---|---|---|
| Air | 1.127 | 1006.4 | 0.02735 | $1.907 \times 10^{-5}$ |
| Aluminum | 2700 | 900 | 238 | - |

**Shape Optimization Approach**

While the effectiveness ($\varepsilon$) is a measure of the heat transfer efficiency between the hot and cold fluids, the pressure-drop ($\Delta P$) is a measure of the operating cost of the heat exchanger. We intend to determine the set of optimum surface parameters, $A$, $B$, $n$, $m$ and $k$ defining the heat transfer surface area that yields the maximum effectiveness while constraining the pressure drop across the heat exchanger within acceptable limits for the same envelope dimension of the reference heat exchanger being the gyroid structure. Additionally, to ensure objective comparison between both architectures in terms of weight consideration, the hydraulic diameter is constrained within range similar to the reference gyroid heat exchanger geometry and similar aluminum material property is assumed for the fin structure. To achieve this, we develop an analytical optimization algorithm in MATLAB (MathWorks Inc, Natick, Massachusetts) that computes the optimum set of the surface control parameters that yields the maximum effectiveness within a given range of pressure drop and hydraulic diameter constraint limits. The effectiveness, pressure drop and hydraulic diameter are outputs returned from the COMSOL numerical FEA solver used to construct the objective and constraints functions. The coupling between the analytical and numerical solver is achieved via COMSOL's MATLAB Livelink module. Mathematically the optimization problem can be recast as

$$\begin{cases} find: \\ \quad \underline{X} = [A \quad B \quad n \quad m \quad k] \\ minimize: \\ \quad -\varepsilon(\underline{X}) \\ Subject \; to: \\ \quad 0 \leq \underline{X} \leq \underline{X}_{max} \\ \quad \Delta P(\underline{X}) \leq \Delta P_{max} \\ \quad D_{h,min} \leq D_h(\underline{X}) \leq D_{h,max} \end{cases} \quad (21)$$

where $\underline{X}$ is the decision variable containing the set of surface control parameters, $\underline{X}_{max}$ is the upper bound limit of the decision variables, $\Delta P_{max}$ is the specified upper bund limit for a given set of operating conditions (i.e. Re = 1000, $T_h^i = 50^0C$, $T_c^i = 30^0C$), and $D_{h,min}$, and $D_{h,max}$ are the lower and upper bounds of the hydraulic diameter, while all other variables retain their usual meaning. To prevent complete occlusion of the channel, the maximum amplitudes of the primary surface terms is set to be one fourth the channel width (i.e. $A_{max} = 1/4 \, W$) and the maximum amplitude of the secondary surface is set to be half that of the primary surface (i.e. $B_{max} = 1/2 \, A_{max}$). Additionally, to manufacturing ensure feasibility, the parameters $n$, $m$, and $k$, were respectively capped at $n_{max} = 2.0$, $m_{max} = 8$ and $k_{max} = 1$. The lower and upper bound limits of $D_h$ was determined based on the hydraulic diameter for the reference gyroid structure (i.e. $D_h^{gyr} = 2.17 \; cm$), as such the limit were set as $D_{h,min} = 2.0 \; cm$ and $D_{h,max} = 2.2 \; cm$. Likewise, the pressure was bound at $\Delta P_{max} = 5.07 \; Pa$, corresponding to the pressure drop for the reference gyroid structure for the same set of operating conditions as would become evident in later sections. The inbuilt surrogate optimization solver `surrogateopt`, from MATLAB's global optimization toolbox, was used to carry out the optimization operation which is suitable for time-consuming objective functions as with our numerical

FEA solver which can be considered as a non-smooth function. The initial values of the decision variables were assumed to be $\underline{X}_0 = [0.25 \quad 0.2 \quad 0.5 \quad 4 \quad 0.01]$. A constraint tolerance of 0.01 was assumed for the linear and non-linear inequality constraints and the maximum number of function evaluations was set to 250.

## Results & Discussion

The results from the FEA simulation based on the design approach are summarized in the following sections. Firstly, the result from the shape optimization studies involving the determination of the optimum second-order harmonic heat transfer surface that yields maximum effectiveness within a specified pressure-drop limit and based on a single set of laminar operating conditions is presented. Additionally, a parameter variation sensitivity study that investigates the unique contribution of the secondary surface tessellations to the HX performance assessed on the basis of the effectiveness and pressure drop by varying the secondary surface control parameters is presented in subsequent sub-sections. Finally the section is concluded with detailed behavioral maps that present a comprehensive overview of the performance of the optimized tessellated wavy channel HX structure in reference to that of the geometric and hydraulically equivalent gyroid TPMS structure considering a broad spectrum of laminar and turbulent Reynolds number for different inlet temperature combinations and assessed on the basis of the HX effectiveness, flow resistance, heat transfer efficiency and overall thermal-hydraulic performance.

**Shape Optimization Studies**

The progression of the optimization states appears in Figure 5 below. In the *'construct-surrogate phase'*, quasi random samples of the objective function (-$\varepsilon$) within the problem bounds are computed by the solver which appear as pink dots. Subsequently in the *'search for minimum phase'*, the solver utilizes a surrogate model to improve the current solution with an adaptive search *'scale'* by determining additional sets of local optimum points around the current solution based on a *'merit function'* while searching for global minimum in unexplored regions. Infeasible adaptive sample points that violate the constraint bounds are denoted by the orange square markers while the feasible points are denoted by the blue dots. The infeasible incumbent points since the previous *'surrogate reset'* are represented by black 'x' markers while the feasible incumbent points are represented by cyan pentagon markers. The 'best' points with the minimum objective function values are represented by circle markers which are orange in color for infeasible points and red in color for feasible points. The vertical lines indicate a *'surrogate reset'* where the solver returns to the *'construct-surrogate phase'* to prevent a local optimum domiciliation. The solver terminates the search when the preset maximum number of function evaluations is reached. From Figure 5, the optimum effectiveness is attained only after about 44 function evaluations which does not change up to 250 function evaluations which indicates successful determination of the global maximum effectiveness for the given set of operating conditions (i.e. Re = $1000, T_h^i = 50^\circ C, T_c^i = 30^\circ C$).

The initial and final geometry from the shape optimization analysis based on the surrogate optimization approach is shown in Figure 6 (a) and (b) respectively. The resulting set of optimum surface control parameters are: $A_{opt} = 0.3898$ cm, $B_{opt} = 0.2232$ cm, $n_{opt} = 1.7464$, $m_{opt} = 6.8285$, and $k_{opt} = 0.5630$. The corresponding hydraulic diameter for the optimized geometry is computed as $D_h = 2.0445$ which satisfies the preset bounds. The optimum effectiveness and associated pressure drop for the given operating conditions are respectively $\varepsilon = 13.9413$ and $\Delta P = 0.41624$ Pa which is far below the preset upper limit (i.e. $\Delta P_{max} = 5.07\ Pa$). The obtained optimum configuration would serve as the basis for comparing the performance of the tessellated wavy channel HX unit cell architecture with the gyroid unit cell structure for various operating conditions which is to follow in subsequent sections.

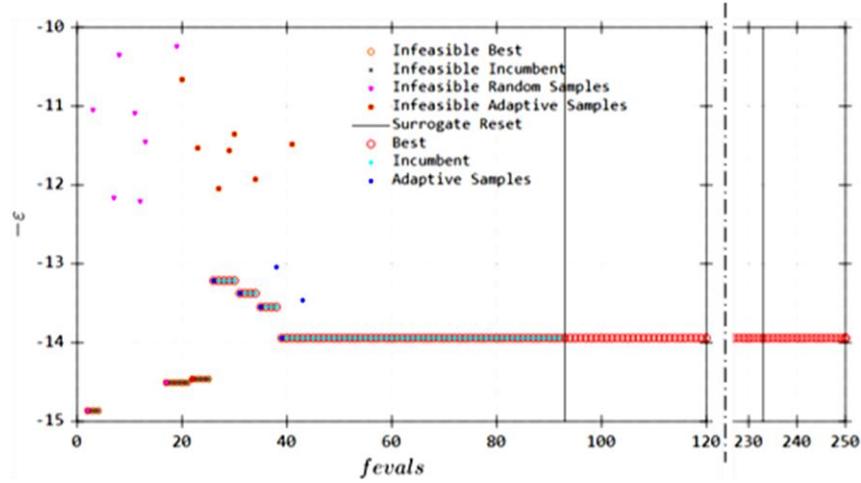

**Figure 5:** Progression of the surrogate optimization algorithm in determining the optimum effectiveness for the given operating conditions (Re $= 1000, \mathcal{T}_h^i = 50^0C, \mathcal{T}_c^i = 30^0C$).

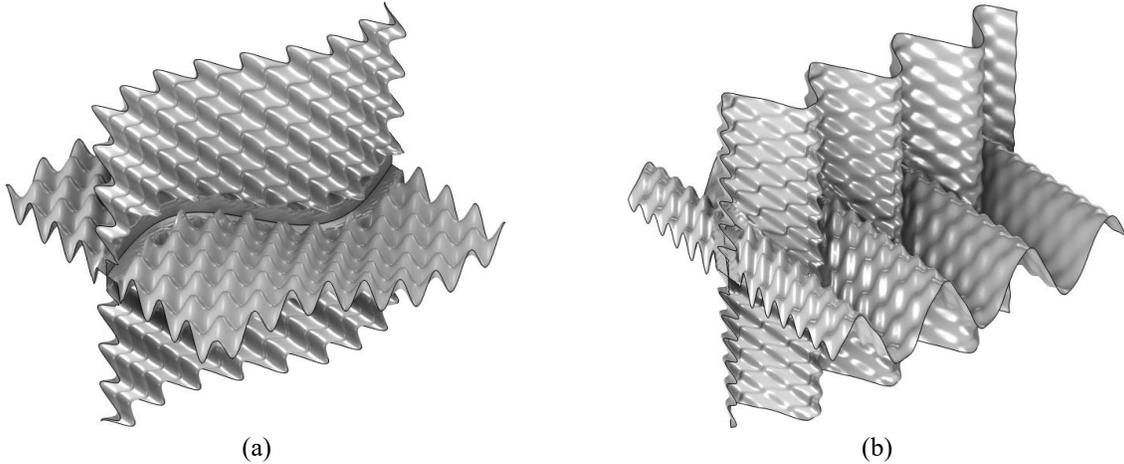

(a)          (b)

**Figure 6:** (a) Initial, and (b) final optimized interior surface geometry of the tessellated wavy channel HX.

**Heat Exchanger Performance Parametric Studies**

*1. Effect of Secondary Surface Parameters:*

To understand the contribution of the secondary surface features to the heat exchanger's performance, a parametric study was carried out that investigates the sensitivity of the heat exchanger's effectiveness, and pressure drop to the parameters defining the secondary features (i.e. wave amplitude – $B$, and wave frequency $m$). Results of the heat exchangers performance for (a) five (5) different amplitudes in the range $0.05\text{cm} \leq B \leq 0.25\text{cm}$ with a corresponding frequency of m $= 4$, and (b) five different frequencies in the range $0 \leq m \leq 8$ with a corresponding amplitude of B $= 0.25\text{cm}$ appear in Figure 7 (a) and (b) respectively.

From Figure 7 (a) and (b), we see what is otherwise expected that increasing the secondary surface control parameters (B and m) results in higher heat exchanger's effectiveness, $\varepsilon$ which is a trade off with a corresponding increase in the pressure drop per unit length, $\Delta P/L$ across the channel. This can directly be attributed to the corresponding increase in the effective heat transfer surface area to volume ratio per channel, $\beta$ whilst the effective hydraulic diameter $D_h$ is decreasing (cf. Figure 7 (c) & (d)). For the given set of operating conditions (i.e. Re $= 3000, \mathcal{T}_h^i = 50^0C, \mathcal{T}_c^i = 30^0C$), the effectiveness varies directly with the surface control parameters ($B$ and $m$) and their relationships can be described by 2nd order polynomials respectively given as

$$\varepsilon(B)|_{m=4} = 21.726B^2 + 5.7371B + 3.0941 \qquad (22)$$
$$\varepsilon(m)|_{B=0.25} = -0.0563m^2 + 1.0189m + 2.6789 \qquad (23)$$

On the other hand, while the pressure drop varies linearly with amplitude $B$ for a constant $m$ (i.e. $m = 4$) with a slope of $55972\ Pa/m^2$, its variation with the wave frequency $m$ for a constant $B$ (i.e. $B = 0.25cm$) can be described by a polynomial given as

$$\left.\frac{\Delta P}{L}(m)\right|_{B=0.25} = 0.6214m^3 - 12.264m^2 + 77.032m + 24.339 \qquad (24)$$

Beyond $m = 4$, the incremental gain in the pressure drop per length declines with increasing $m$ and reaches a maximum when $m \approx 5$ while the effectiveness continuously increases with $m$.

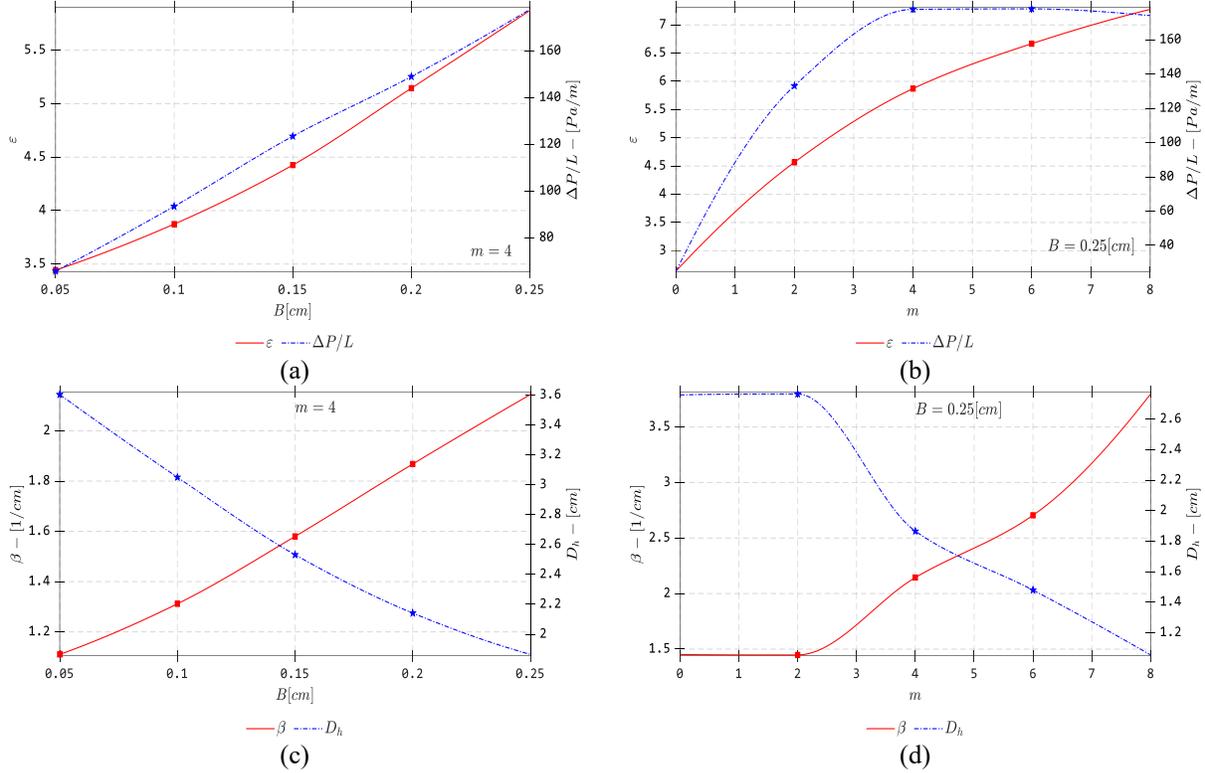

**Figure 7:** Results of the effectiveness, $\varepsilon$ (red line) and specific pressure drop, $\Delta P/L$ (blue line) for different values of (a) amplitudes, $B$, and (b) frequency, $m$. Results of the surface area density, $\beta$ (red line) and hydraulic diameter, $D_h$ (blue line) for different values of (c) amplitudes, $B$, and (d) frequency.

Besides the fact that increasing the magnitudes of the control parameters results in higher heat transfer surface area to volume ratio per channel, $\beta$, the associated channel surface roughness due to the surface tessellations induces local vorticity near the channel walls and also results in higher velocity magnitudes (cf. Figure 8). This leads to an increase in the heat exchange via convection and an overall increase in the heat exchanger's effectiveness. Although the heat exchanger's performance is seen to directly relate to the surface control parameters, it is however constrained by the feasibility of manufacturing such intricate details. Another concern with this type of heat transfer surface is its susceptibility to fouling which introduces some degree of thermal resistance. As such, the fluid and operating conditions need to be taken into consideration in the design section.

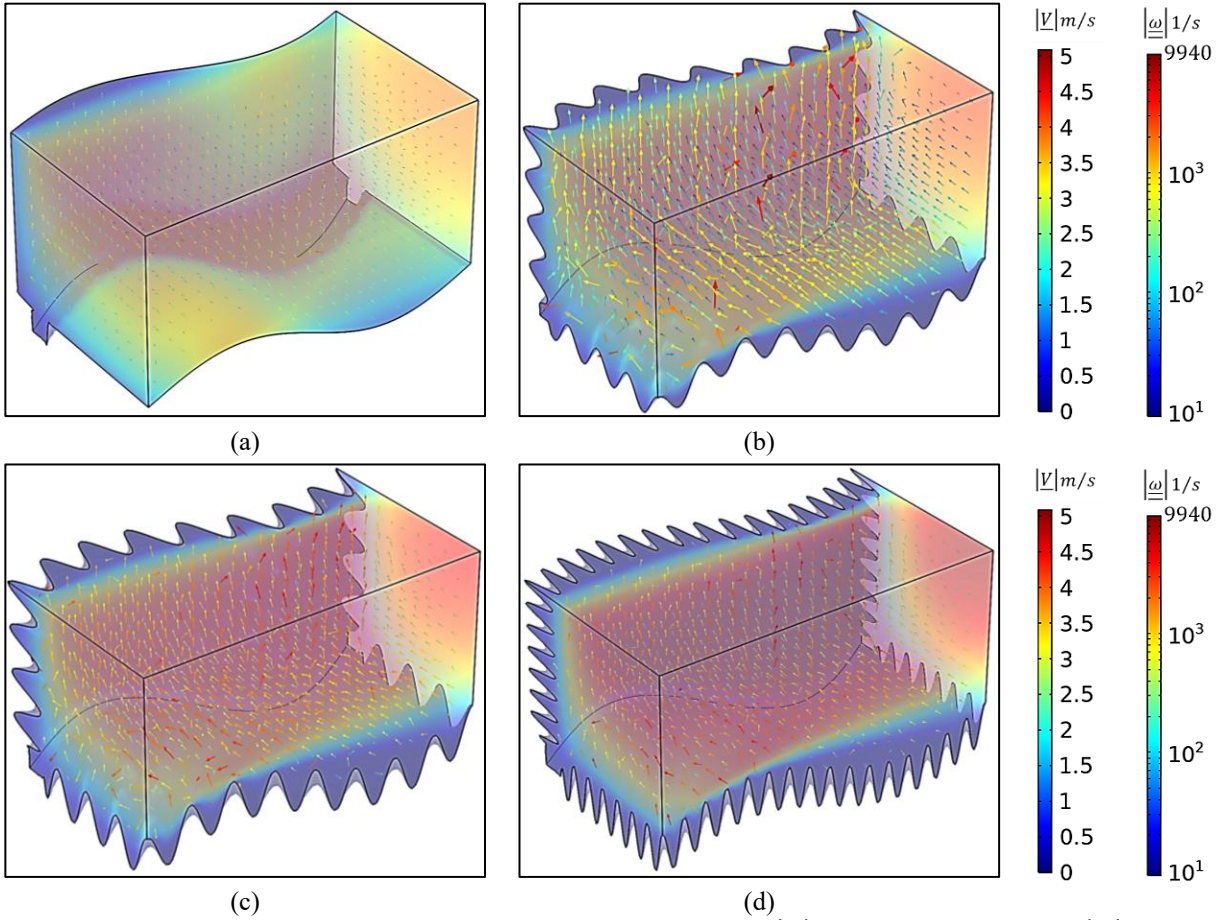

**Figure 8:** Contour plot showing the distribution of the velocity magnitude, $|\underline{V}|$, and vorticity magnitude, $|\underline{\omega}|$, (quivers) in a single tessellated wavy channel, for different heat transfer surface configurations (a) smooth channel, m=0 (b) B=0.15cm, m=4, (c) B=0.25cm, m=4, and (d) B=0.25cm, m=8.

## 2. Effect of Operating Conditions:

We construct a performance map to understand the behavior of the proposed heat exchanger design under a given set of inlet temperatures, $T_h^i = 40, 50, 60\ °C$, $T_c^i = 20, 30\ °C$ and for various flow regimes with Reynold's number within the range $200 \leq Re \leq 10,000$ and compare results between the tessellated wavy channel structure and the gyroid architecture. Figure xx (a) and (b) shows error-bar plots for the resulting effectiveness, specific pressure-drop and LMTD profiles as a function of the Reynold's number for the various operating conditions. Overall, we observe relatively decreasing effectiveness with increasing Reynolds number over the range of operating temperatures considered for both tessellated wavy and gyroid HX architectures (cf. Figure 9(a)). In the laminar regime, the effectiveness of the gyroid structure is seen to be relatively higher compared to the tessellated wavy HX structure, however the relative difference in the effectiveness is seen to decrease gradually with increasing Reynolds number at a rate of $\Delta \varepsilon = 52.77 Re^{-0.1072} - 20.55$, until a crossing point at $Re = 7000$, where the effectiveness of the tessellated wavy structure surpasses that of the gyroid structure. Although higher Reynold's number increases the rate of convective heat transfer and promotes turbulent mixing, however, the corresponding increase in the effective mass flowrate reduces the fluid residence time in the channel and the resulting inlet-outlet temperature difference. The HX effectiveness thus depends on the competing influence of the convective heat transfer rate and fluid residence time. On the other hand, the specific pressure-drop increases with increasing Reynold's number and the gyroid structure is observed to have relative higher pressure drop compared to tessellated wavy channel structure which is on average about 10.52 times higher with a standard deviation of 1.61 in the laminar regime, i.e. $Re < 2000$, and about 2.75 times higher with a standard deviation of 0.1 in the turbulent regime, i.e. $Re \geq 3000$, (cf. Figure 9(b)). Despite having similar specific heat transfer surface area, the relatively higher pressure-drop in the gyroid structure compared to that of the tessellated wavy structure can be attributed to its tortuous flow path which significantly increases the form (pressure) drag in comparison to the skin friction drag. One can develop approximate expressions that describe the relationship between the effectiveness, $\varepsilon$ and specific pressure drop, $\Delta P/L$ for both designs via a nonlinear regression fitting procedure. For the tessellated wavy channel architecture, the approximate correlations for $\varepsilon$ and $\Delta P/L$ with respect to $Re$ are respectively given as

$$\varepsilon_{tss} \approx 15.41 e^{-5.685 \times 10^{-4} Re} + 5.145 e^{2.384 \times 10^{-5} Re} \tag{25}$$

$$\frac{\Delta P_{tss}}{L} \approx 6.254 \times 10^{-5} Re^{1.967} \, [Pa \cdot m^{-1}] \tag{26}$$

Likewise, for the gyroid unit cell, the approximate correlations for $\varepsilon$ and $\Delta P/L$ with respect to $Re$ are respectively given as

$$\varepsilon_{gyr} \approx 24.31 e^{-6.792 \times 10^{-4} Re} + 6.692 e^{-9.876 \times 10^{-6} Re} \tag{27}$$

$$\frac{\Delta P_{gyr}}{L} \approx 3.872 \times 10^{-4} Re^{1.874} \, [Pa \cdot m^{-1}] \tag{28}$$

where subscript '$tss$' indicates parameters for the tessellated wavy channel HX architecture, and subscript '$gyr$' indicates parameters for the gyroid HX structure.

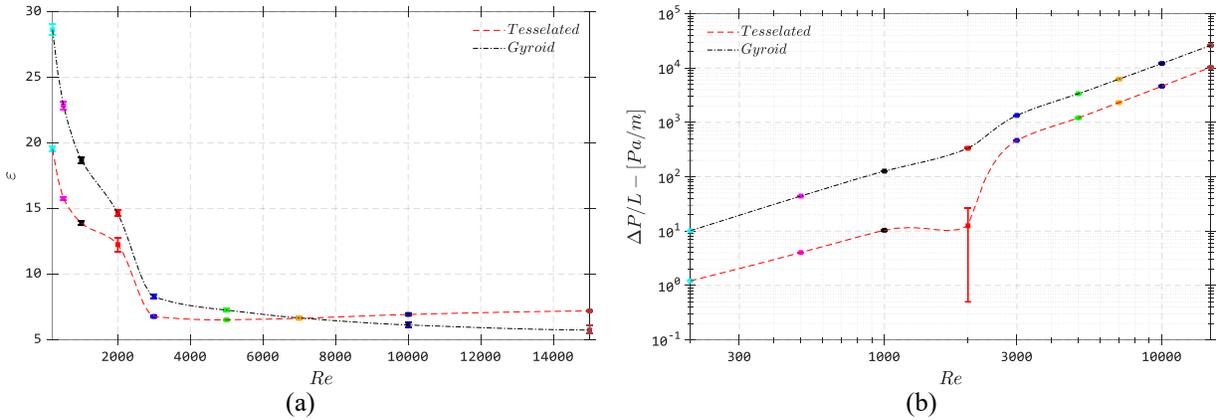

**Figure 9:** Errorbar plots of the heat exchanegrs (a) effectiveness, $\varepsilon$ (b) specific pressure drop, $\Delta P/L$ versus the Reynold's number, $Re$ for both tesselated wavy channel stucture (red line) and gyroid structure (black line).

The LMTD, $\Delta T_m$ is a measure of the average temperature difference between both hot and cold fluids. A higher $\Delta T_m$ is desirable since this indicates a lesser heat transfer surface area is required to achieve the same heat transfer rate in both HX architectures. From Figure 10 below, the LMTD, $\Delta T_m$ is seen to be on average relatively higher in the laminar regime for the tessellated wavy channel architecture compared to the gyroid structure. Although we observe an anomaly around $Re = 2000$ for tessellated wave architecture which is likely due to instability that characterize the transition flow regime. As the Reynold's number increases and the flow transitions to fully turbulent condition, $\Delta T_m$ is seen to increase however at a decaying rate, almost flattening out at very high Reynold's number ($Re > 5000$) beyond which both HX structures begin to have almost matching $\Delta T_m$ profiles. The variation in the LMTD limits and interquartile ranges for the different inlet temperature combinations likewise increases with increasing Reynold's number until a point where it tends to stabilize (at about $Re > 5000$).

Velocity contour plot superimposed on vorticity quiver plots for the reference inlet temperature conditions (i.e. $T_h^i = 50^0C$, $T_c^i = 30^0C$), and for select flow regimes (i.e. $Re = 3000, 5000, 7000 \, and \, 10,000$) are shown in Figure 11(a)-(d) for the tessellated wavy channel HX structure and in Figure 12(a) – (d) for the gyroid HX architecture. As expected, the average velocity magnitude increases with increased Reynolds number due to increased mass flux through the channel. The velocity magnitude decreases towards the walls of the channel which is almost zero within troughs of the primary surface of the tessellated wavy channel HX structure irrespective of the Reynolds number. Away from the channel walls, the velocity distribution is observed to be nearly uniform. On the contrary, the velocity distribution within the gyroid flow domain is characterized by regions of flow acceleration and deceleration due to the tortuous flow-path that creates stagnation zones. Overall, the distribution of the average velocity in the gyroid unit cell fluctuates over a wider range, e.g. from 0 to $34.1\,m/s$ for $Re = 10{,}000$, which is about 1.7 times more than the value observed in the tessellated wavy channel structure. While the associated vorticity tensor magnitude also increases with increasing Reynolds number, it is however observed to be more uniformly distributed within the flow domain of the gyroid unit cell compared to the tessellated wavy channel structure where it is seen to be dominant closer to the channel walls. Despite the gyroid unit cell having relatively higher velocities due to the tortuosity of the flow-path, interestingly, the distribution of the vorticity magnitude is seen to be relatively lower, for instance, approximately 2.25 times less than that observed in the tessellated wavy channel HX counterpart for $Re = 10{,}000$. This can be attributed to the high degree of surface undulations that characterize the channel walls of the tessellated structure which induces local vortices near the channel walls that promote increased level of convective heat transfer between the wall and fluid. These secondary flows may thin or disrupt the thermal boundary layer to enhance fluid mixing.

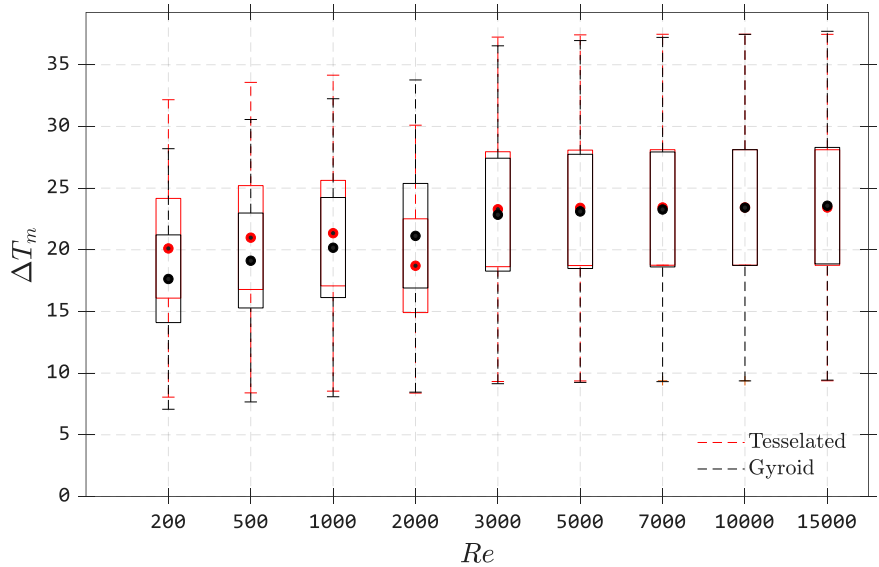

**Figure 10:** Boxplot showing the mean values (inner dots), interquartile intervals (box limits), and extremums (whiskers) of the computed LMTD for the various inlet temperature combinations, Reynold's number and for both HX architectures.

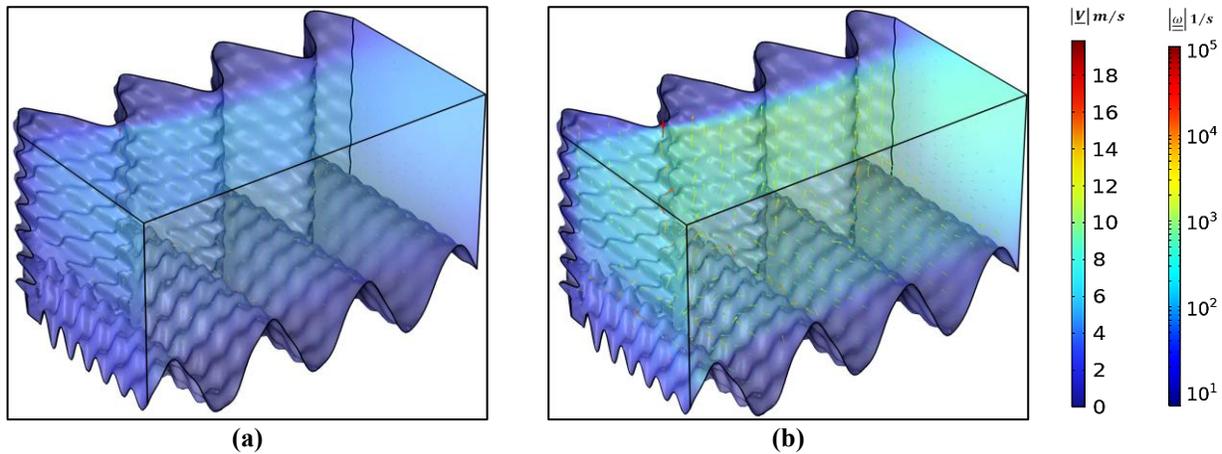

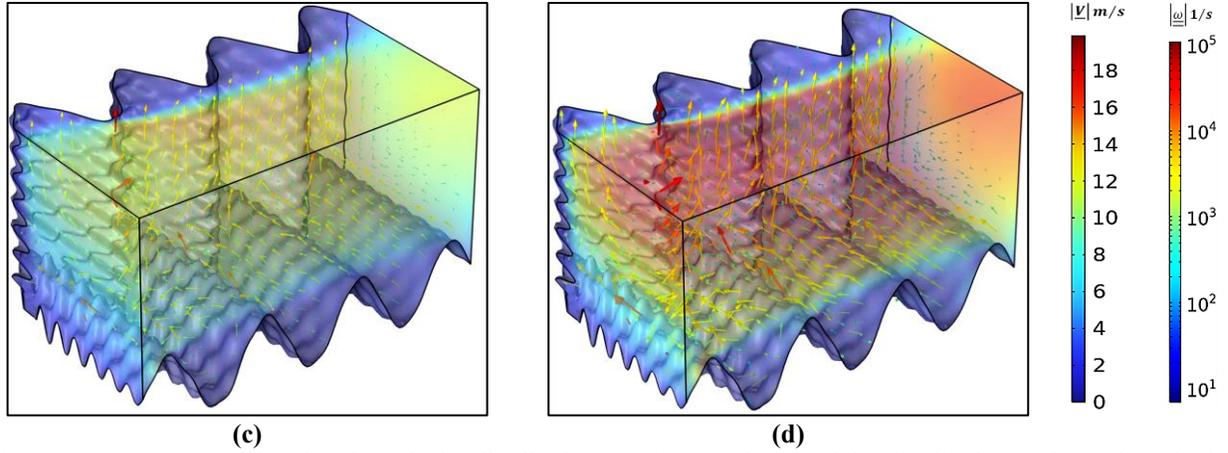

**Figure 11:** Contour plot showing the velocity distribution superimposed on vorticity distribution (quivers) in a single channel of the tessellated wavy structure for different turbulent Reynold's number (a) $Re = 3000$, (b) $Re = 5000$, (c) $Re = 7000$, and (d) $Re = 10,000$, for inlet temperatures $T_h^i = 50^0C$, $T_c^i = 30^0C$.

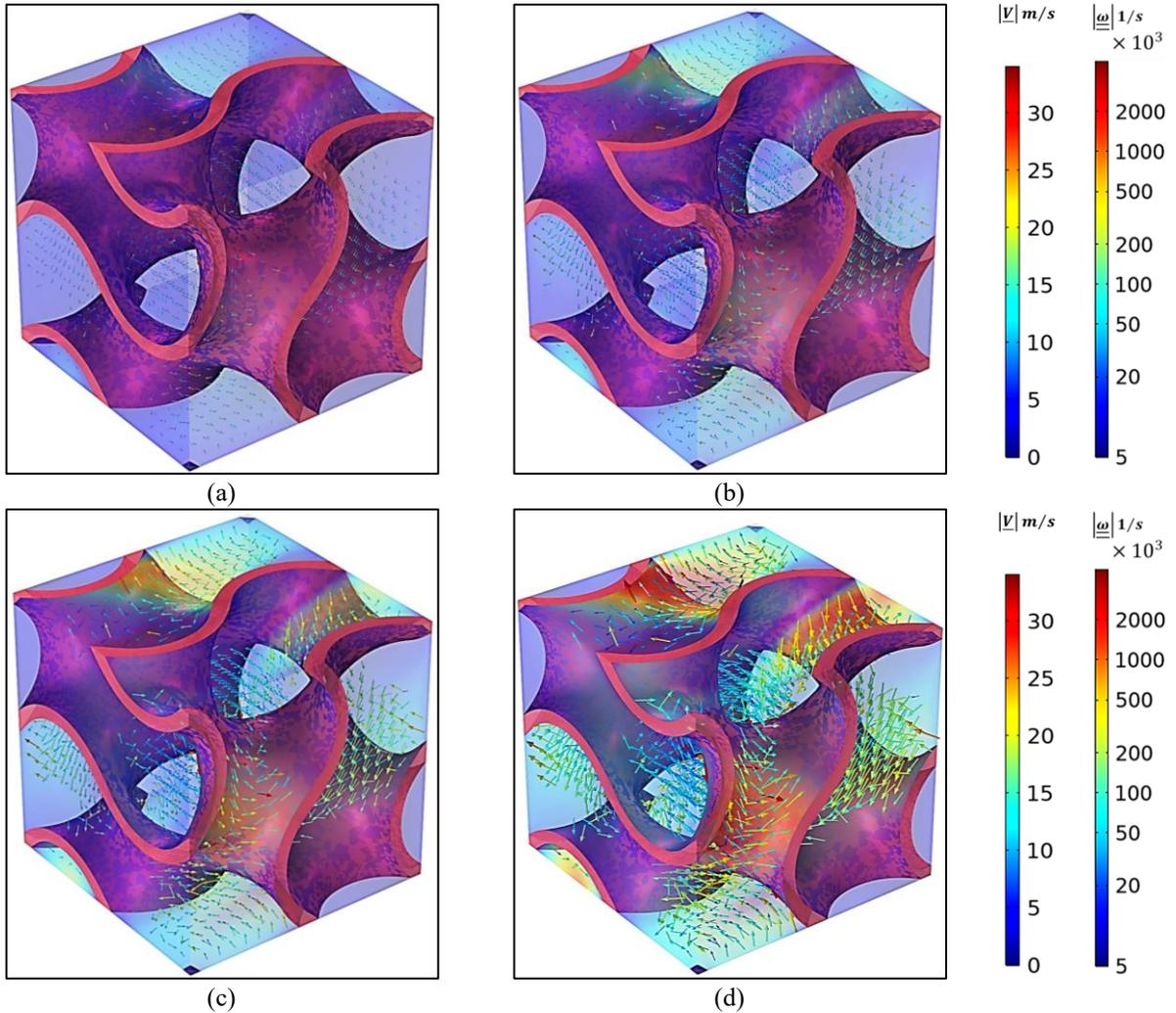

**Figure 12:** Contour plot showing the velocity distribution superimposed on vorticity distribution (quivers) within the gyroid unit cell flow domain for different turbulent Reynold's number (a) $Re = 3000$, (b) $Re = 5000$, (c) $Re = 7000$, and (d) $Re = 10,000$, for inlet temperatures $T_h^i = 50^0C$, $T_c^i = 30^0C$.

As previously noted, the HX effectiveness was observed to be generally low in the turbulent regime with high pumping requirements based on the pressure drop profiles (cf. Figure 9(b)). On this basis, it may seem desirable to operate the HX in the laminar flow regime although most air-to-air HX in practical applications operate in the turbulent regime. However, a more important parameter used to quantify the HX heat transfer performance the Nusselt number, $Nu$ evaluated using eqn. xx. The computed Nusselt number for the different inlet temperature combinations and Reynolds number are shown in the error-bar plot of Figure 13(a) for both gyroid and tessellated wavy channel architecture. The Nusselt number abstracts the operating temperatures and fluid type from the quantification of the heat transfer efficiency. On this basis, the average Nusselt number is observed to increase almost linearly with increasing Reynolds number which indicates an overall increase in the heat transfer coefficients in both architectures. Linear correlation trends based on regression fitting of the computed datapoints to the Seider-Tate type equation (cf. eqn. (11)) appear in Figure 13(a) for the tessellated wavy channel structure (red dashed lined) and the gyroid structure (black dashed line) which are mathematically represented by eqn. (29) and (30) respectively.

$$Nu_{tss} = -4.11 + 1.04 \left( Pr \times Re \times \frac{D_h}{l_c} \right)^{1/3} \tag{29}$$

$$Nu_{gyr} = -21.45 + 5.31 \left( Pr \times Re \times \frac{D_h}{l_c} \right)^{1/3} \tag{30}$$

Evidently, the computed Nusselt number for the gyroid structure, $Nu_{gyr}$ is much higher (approximately 5.1 times greater) than that of the tessellated wavy channel structure, $Nu_{tss}$ which implies a higher heat transfer efficiency. This expected superior heat transfer performance however comes at the cost of relatively higher pressure drop which is quantified by the dimensionless Fanning friction factor, $f$ (cf. eqn. (14)) as shown in Figure 13(b). Explicit correlations for the friction factor $f$, for the tessellated wavy channel structure (red dashed line) and the gyroid structure (black dashed line) derived from non-linear regression analysis based on the form of eqn. (14) [22] are respectively presented in eqns. (31) and (32) below.

$$f_{tss} \cdot Re = \begin{cases} 17.68 + 0.00162 Re^{1.323}, & Re \leq 2000 \\ 234.3 + 0.14320 Re^{1.020}, & Re \geq 3000 \end{cases} \tag{31}$$

$$f_{gyr} \cdot Re = \begin{cases} -24.68 + 6.166 Re^{0.585}, & Re \leq 2000 \\ 206.1 + 0.653 Re^{0.932}, & Re \geq 3000 \end{cases} \tag{32}$$

Although, alternative expressions for both architectures could also be derived based on second order polynomial regression fitting of $f$ to the natural logarithm of $Re$ which are respectively given as

$$f_{tss} = \begin{cases} 0.897 - 0.2356 \ln Re + 0.016 (\ln Re)^2, & Re \leq 2000 \\ 1.952 - 0.3613 \ln Re + 0.019 (\ln Re)^2, & Re > 2000 \end{cases} \tag{33}$$

$$f_{gyr} = \begin{cases} 1.894 - 0.3341 \ln Re + 0.01548 (\ln Re)^2, & Re \leq 2000 \\ 2.239 - 0.3616 \ln Re + 0.01723 (\ln Re)^2, & Re > 2000 \end{cases} \tag{34}$$

The above expressions characterize two flow regimes, i.e. the laminar flow regime ($Re \leq 2000$), and the turbulent regime ($Re \geq 3000$). In either case, the friction factor is significantly higher for the gyroid structure in comparison to the tessellated wavy channel structure which invariably implies a higher pumping requirement. The suitability and choice of a design in the context of performance is thus application specific and would depend on the appropriation of weighting factors to the various performance metrics (i.e. heat transfer efficiency parameter, $Nu$, and pumping requirement parameter, $f$) depending on their individual level of importance. In terms of the effectiveness, if the heat exchanger were to operate in the laminar and the weakly turbulent flow regime, the suitability of design choice would depend on associated pressure drop, however in the fully turbulent region beyond $Re > 7000$, the tessellated wavy channel structure with relatively higher effectiveness and lower pressure drop would apparently outperform the gyroid structure. Based on the heat transfer efficiency parameter ($Nu$) however, the gyroid structure although appears to be preferable, its seemingly superior performance is a trade-off with the frictional resistance. The later argument creates a premise for the establishment of a so-called '*overall*' thermal-hydraulic performance metric to assess the fitness of purpose of a design selection. One commonly used metric is the well-known London area goodness factor - $j/f$ [26,28] which is an evaluation index that gives a comprehensive measure of the heat transfer performance with respect to the pressure drop for fixed flow area. Figure 14 shows the variation of the computed $j/f$ with Reynold's number, $Re$, for the different inlet temperature combinations and for the tessellated wavy channel HX structure (red line) and the gyroid structure (black line). Counterintuitively, the metric shows that the tessellated wavy channel structure has superior overall performance in comparison to the gyroid structure in the laminar flow regime ($Re \leq 2000$), reaching up to about 2.2 times the AGF of the gyroid structure at about $Re = 500$, whereas its computed effectiveness is relatively lower compared to gyroid structure (cf. Figure 9(a)). In the turbulent flow regime ($Re \geq 3000$), the tessellated wavy channel structure is seen to have relatively lower AGF or $j/f$, on average about 0.35 times that of the gyroid structure, however one can argue that the latter has relatively higher effectiveness and lower pressure drop in this region. Moreover, the HX effectiveness in the turbulent regime is generally low for both architectures.

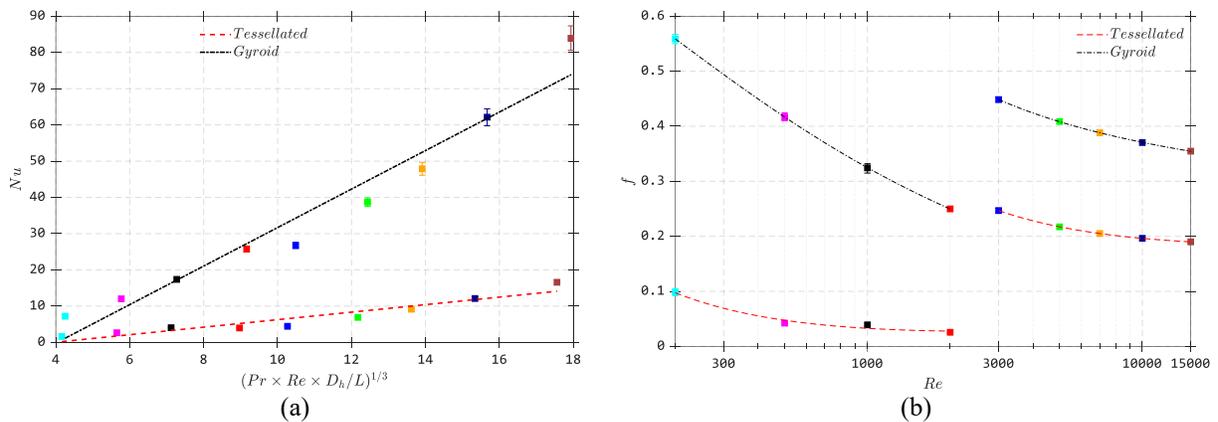

**Figure 13:** Error-bar plot showing the variation of the (a) Nusselt number, $Nu$, and (b) friction factor, $f$, with the Reynolds numbers for different inlet temperature combinations. Also shown are the trend lines from regression fitting procedures for the tessellated wavy channel structure (black lines), and gyroid structure (red lines).

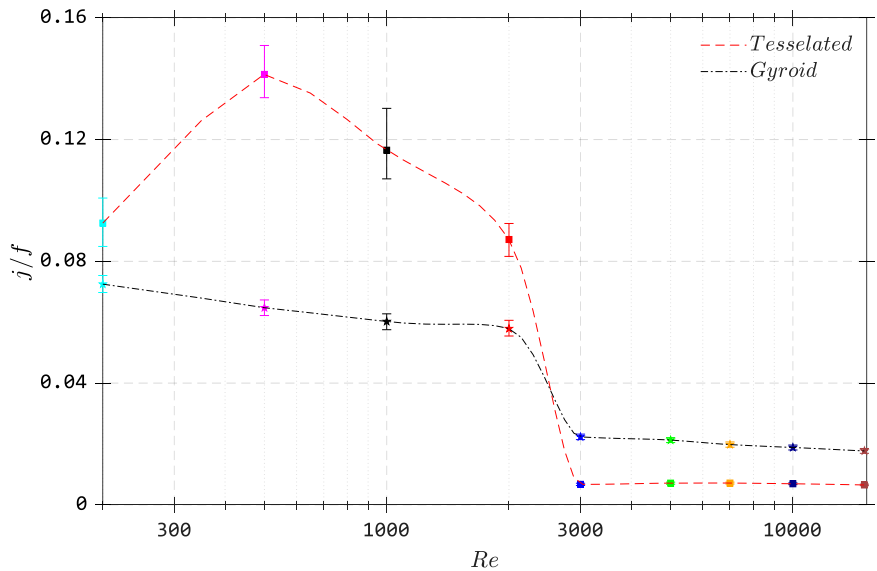

**Figure 14:** Error-bar plot showing the variation of the computed "London area goodness factor" (AGF) - $j/f$ with Reynold's number, $Re$, for different inlet temperature combinations; for the tessellated wavy channel HX structure (red line) and the gyroid structure (black line).

From the foregoing discussion, it can be concluded that while the gyroid TPMS structure may offer superior heat transfer efficiency owing to its topological nature, the design suffers from high levels of associated pressure drop that results from the complicated and highly convoluted flow-path and its overall performance is subject to interpretation. With the continuous advancement and growth of additive manufacturing technology, heat transfer enhancement via surface modification with unique features can be achieved with comparable HX effectiveness as TPMS structures and without necessarily incurring high degree of pressure drop. The optimization of second-order harmonic type heat transfer surfaces investigated in this work as a proof-of-concept design has revealed higher AGF or $j/f$ in the laminar flow region with reasonable HX effectiveness, however the structure loses its advantage in the turbulent flow regime. Yet the research reveals opportunities for utilization of exotic surface equations other than that investigated in this work for design specific and fit for purpose application. Other design factors such as fouling, design feasibility, manufacturing cost, serviceability, etc. would ideally be taken into account for a more comprehensive and broader scale assessment of the overall design advantage.

## Conclusion

In conclusion, the current research has successfully investigated the performance of a heat exchanger unit cell having heat transfer surfaces described by second-order harmonic type tessellations, in comparison to a geometrical and hydraulically equivalent gyroid TPMS unit cell. The comparison is made on the basis of several performance metrics including the effectiveness, pressure-drop, heat transfer efficiency, flow resistance, and overall thermal-hydraulic performance.

Preliminary sensitivity studies that investigate the unique contribution of the secondary surface features to the heat exchangers performance were carried out by varying the secondary surface control parameters. It was observed that the heat exchangers effectiveness increased at a rate of $\partial\varepsilon/\partial B = 21.726B + 5.7371$, by varying the secondary surface amplitude $B$, and at a rate of $\partial\varepsilon/\partial m = -0.0563m + 1.0189$ by varying the frequency $m$. However, the observed improvement in the heat exchangers effectiveness by secondary surface modifications comes at the expense of an associated increase in the pressure drop. A five (5) times increase in the amplitude, $B$ results in a 70.5% increase in the effectiveness and an associated increase of 170.8% in the specific pressure drop while multiplying the frequency $m$, by four (4) increases the effectiveness by about 59.3% and an associated increase of just 30.8% in the specific pressure drop. This implies that the frequency $m$ is a more important parameter in improving the overall thermal-hydraulic performance of the heat exchanger. The pressure drop profile is seen to flatten out beyond $m > 4$ with a noticeable drop at $m = 8$ whilst the effectiveness continually increases with $m$ at an almost linear rate.

Optimization of the second order harmonic heat transfer surface was also carried out to determine the optimum set of surface control parameters that yielded the maximum effectiveness for a given set of operating conditions corresponding to $Re = 1000, T_h^i = 50°C, T_c^i = 30°C$, within the limit of a preset maximum pressure drop constraint and within a limited range of the hydraulic diameter similar to that of the gyroid structure. The choice to optimize in the laminar flow regime was informed by the fact that the maximum effectiveness occurs in this flow regime and considering only a single optimum geometry can selected when benchmarking its performance with the gyroid TPMS structure for all operating cases. The optimization was achieved by utilizing the MATLAB's inbuilt surrogate optimization solver from the global optimization toolbox which was coupled to COMSOL Multiphysics FEA solver via the MATLAB Livelink module. The resulting set of surface control parameters from the optimization operation were $A_{opt} = 0.3898$ cm, $B_{opt} = 0.2232$ cm, $n_{opt} = 1.7464$, $m_{opt} = 6.8285$, and $k_{opt} = 0.5630$ corresponding to a hydraulic diameter of $D_{h,opt} = 2.0445$, an effectiveness of $\varepsilon_{opt} = 13.9413$ which improve by about 13.62% from the initial value, and a pressure drop of $\Delta P = 0.41624$ Pa which was well below the set maximum value of $\Delta P_{max} = 5.07\ Pa$.

Ultimately the performance of the optimized tessellated wavy channel structure is compared to that of the gyroid structure using various metrics. The gyroid structure had relatively higher effectiveness mostly in the laminar flow regime compared to that of the tessellated wavy channel structure although the effectiveness ratio ($\varepsilon_{\text{gyr}}/\varepsilon_{tss}$) declined steadily with increasing Reynold's number from about $\varepsilon_{\text{gyr}}/\varepsilon_{tss} = 1.46$ when $Re = 200$ to unity when $Re = 7000$. Beyond, $Re > 7000$, the effectiveness ratio fell steadily below unity to about $\varepsilon_{\text{gyr}}/\varepsilon_{tss} = 0.8$ when $Re = 15000$. However, the associated specific pressure drop of the gyroid structure was significantly higher in all flow regimes, on average about 14.55 times higher in the laminar flow regime (i.e. $Re \leq 2000$) and about 2.75 times higher in the turbulent flow regime (i.e. $Re \geq 3000$). On this basis, the tessellated wavy channel structure outperforms the gyroid structure when $Re \geq 7000$, however the relative performance is subject to interpretation when $Re < 7000$. Based on the heat transfer efficiency quantified by the Nusselt number, $Nu$, the gyroid structure was found to have relatively higher values, on average about 5.2 times that of the tessellated wavy channel structure over the entire range of Reynold's number consider. However, the flow resistance of the gyroid structure quantified by the Fanning friction factor was also found to be significantly higher than that of the tessellated wavy channel structure, on average about 8.4 times higher in the laminar flow regime (i.e. $Re \leq 2000$) and about 1.87 times higher in the turbulent flow regime (i.e. $Re \geq 3000$). The overall thermal-hydraulic performance of both structures was assessed based on the AGF or $j/f$ ratio which revealed a higher performance for the tessellated wavy channel structure compared to the gyroid structure in the laminar flow regime (i.e. $Re \leq 2000$) and vice versa in the turbulent regime.

In summary, the suitability of a particular HX design would depend on its specific application and the desired operating conditions. The structure could be designed either for high heat transfer efficiency with significant pumping requirement or for moderate heat transfer efficiency with low pumping requirement. Likewise, the structure could be designed to possess superior performance in the laminar flow regime or in the turbulent flow regime. Irrespective of the design requirements, the desired performance could be achieved via additively manufacturing of fit-for-purpose complex structures with engineered surface topology.

**CRediT authorship contribution statement**
**Patrick Adegbaye:** Conceptualization, Methodology, Investigation, Data curation, Formal analysis, Visualization, Validation, Writing – original draft, Writing – review & editing. **Aigbe Awenlimobor:** Conceptualization, Methodology, Investigation, Data curation, Formal analysis, Visualization, Validation, Writing – original draft, Writing – review & editing. **Justin An:** Writing – review & editing. **Zhang Xiao:** Validation, Writing – review & editing; **Jiajun Xu:** Conceptualization, Methodology, Validation, Writing – review & editing, Software, Resources, Project administration, Funding acquisition.

**Declaration of Competing Interest**
The authors declare that they have no known competing financial interests or personal relationships that could have appeared to influence the work reported in this paper.

**Acknowledgments**
This project was supported by the U.S. Department of Energy (DOE) Office of Energy Efficiency and Renewable Energy (EERE) - Industrial Technologies Office (ITO), (Award Number: DE-EE0010861).

**Data availability**
Data will be made available on request